\documentclass{emulateapj}
\usepackage{color}
\usepackage{latexsym, graphicx, amssymb, longtable,epsf,ulem}
\usepackage{cases}
\newcommand{\hii}{H{\sc ii~}}
\newcommand{\ant}{\rm {\alpha_{nt}}}

\newcommand{\qlband}{q_{20\rm cm}}
\newcommand{\qpband}{q_{90\rm cm}}

\long\def\symbolfootnote[#1]#2{\begingroup\def\thefootnote{\fnsymbol{footnote}}
\footnote[#1]{#2}\endgroup}

\shorttitle{Radio--FIR correlation}
\shortauthors{Basu, Roy \& Mitra}

\begin{document}

\title{Low frequency radio--FIR correlation in normal
galaxies at $\sim$ 1 kpc scales}

\author{Aritra Basu, Subhashis Roy and Dipanjan Mitra}
\affil{National Center for Radio Astrophysics, TIFR, Pune University
Campus, Ganeshkhind Road, Pune - 411007 \\ aritra@ncra.tifr.res.in (AB);
roy@ncra.tifr.res.in (SR); dmitra@ncra.tifr.res.in (DM)}


\begin{abstract}

We study the radio--FIR correlation between the nonthermal (synchrotron) radio
continuum emission at $\lambda90$ cm (333 MHz) and the far infrared emission
due to cool ($\sim20$ K) dust at $\lambda70~\mu$m  in spatially resolved normal
galaxies at scales of $\sim$1 kpc.  The slope of the radio--FIR correlation
significantly differs between the arm and interarm regions. However, this
change is not evident at a lower wavelength of $\lambda20$ cm (1.4 GHz).  We
find the slope of the correlation in the arm to be $0.8\pm0.12$ and we use this
to determine the coupling between equipartition magnetic field ($B_{\rm eq}$)
and gas density ($\rho_{\rm gas}$) as $B_{\rm eq} \propto \rho_{\rm gas}^{0.51
\pm 0.12}$. This is close to what is predicted by MHD simulations of turbulent
ISM, provided the same region produces both the radio and far infrared
emission. We argue that at 1 kpc scales this condition is satisfied for radio
emission at 1.4 GHz and may not be satisfied at 333 MHz. Change of slope
observed in the interarm region could be caused by propagation of low energy
($\sim$ 1.5 GeV) and long lived ($\sim10^8$ yr) cosmic ray electrons at 333
MHz.

\end{abstract}



\section{Introduction}
\label{intro}

The radio--far infrared (FIR) correlation in normal galaxies was first observed
by \cite{kruit71, kruit73} and later extended by the IRAS mission. Subsequently
it was established that the correlation holds good (within a factor of 2) over
five orders of magnitude in radio and FIR luminosity \citep{condo92,yun01} for
a wide morphological class of galaxies like, spirals, irregulars and dwarfs
\citep{wunde87,dress88,price92} on global scales. Based on spatially resolved
studies of normal and irregular galaxies it is seen that the correlation holds
even at scales of few tens to hundreds of parsecs \citep[see e.g][]{beck88,
xu92, hoern98, murgi05, tabat07a, hughe06, murph06a, palad06, palad09, dumas11}. 

The basic model that connects these two regimes of emission is via star
formation \citep{harwi75}. The radio continuum emission arises due to
synchrotron emission (henceforth nonthermal emission) from relativistic
electrons, produced in supernova remnants. A good fraction of them originate
from massive ($\gtrsim 10~ \rm M_\odot$), short lived ($\lesssim10^6$ yr)
stars. The FIR emission arises from re-radiation by dust heated due to ultra
violet (UV) photons emitted by the above population of stars.  Though the cause
of the correlation is well understood, the tightness over several orders of
magnitude still remains puzzling. Many models explaining the correlation
require close coupling between the magnetic field ($B$) and the gas density
($\rho_{\rm gas}$) of the form, $B\propto\rho_{\rm gas}^{\kappa}$ \citep[see
e.g.,] []{helou93, nikla97, thomp06}. Such a coupling can be established by
magnetohydrodynamic (MHD) turbulence of the interstellar medium (ISM)
\citep[see][]{chand53, cho00, cho03, grove03}. Numerical simulations by
\citet{cho00} revealed that $\kappa = 0.5$ is a manifestation of the
equipartition condition, i.e, in steady MHD turbulence the magnetic field
energy density and the energy density of the gas are similar. Similar values of
$\kappa$ have been found through observations of magnetic field by Zeeman
splitting observations in molecular clouds by \cite{crutc99}, also by using
equipartition magnetic field and molecular gas observations in external
galaxies by \citet{nikla97} and in Milky Way and M31 by \citet{berkh97}.
Alternatively, the slope of the radio--FIR correlation has been used to find
$\kappa$, where $\kappa\sim$ 0.4--0.6 \citep{nikla97, hoern98, dumas11}.

\begin{table*} 
\begin{centering} 
\caption{The sample galaxies.} 
\begin{tabular}{@{}lcccccccc@{}} 
\hline 
    Name  & Morphological       &Angular           & $i$       & Distance     &  FIR             &  &Radio~~~~~~~~~~~~~~~~~~~~~~~~~~~~~ \\ 
          &type       & size (D$_{25}$)($^\prime$)& ($^\circ$) & (Mpc)      &  $\lambda70\mu$m &$\lambda90$cm&$\lambda20$cm\\ 
     (1)     &(2)       & (3) & (4) & (5)               & (6) & (7) &(8)\\ 
\hline 
NGC 4736    & SAab     & 11.2$\times$9.1  & 41 & 4.66$^1$     &SINGS&GMRT& Westerbork\footnote{The Westerbork
Synthesis Radio Telescope (WSRT) is operated by the Netherlands Foundation for
Research in Astronomy (NFRA) with financial support from the Netherlands
Organization for scientific research (NWO).} SINGS (1374.5 MHz)$^4$      \\ 
NGC 5055    & SAbc     & 12.6$\times$7.2  & 59 & 9.2$^\dagger$&SINGS&GMRT& Westerbork SINGS (1696 MHz)$^4$  \\ 
NGC 5236    & SABc       & 11.2$\times$11      & 24 & 4.51$^2$ &SINGS&GMRT&  VLA\footnote{The Very Large Array (VLA) is operated
by the NRAO. The NRAO is a facility of the National Science Foundation operated
under cooperative agreement by Associated Universities, Inc.} CD array (1452 MHz)$^5$ \\ 
NGC 6946    & SABcd    & 11.5$\times$9.8  & 33 & 6.8$^3$       &SINGS&GMRT& VLA C$+$D array (1465 MHz)$^6$        \\ 
\hline 
\end{tabular}\\ 
In column (3)  D$_{25}$ refers to the optical diameter measured at the 25 
magnitude arcsec$^{-2}$ contour from \cite{vauco91}.  Column (4) gives the 
inclination angle ($i$) defined such that $0^\circ$ is face-on.  Distances in 
column (5) are taken from: $^1$ \cite{karac03}, $^2$ \cite{karac02}, $^3$ 
\cite{karac00} and the NED $^\dagger$. Column (6) and (7) are the sources of 
data for the FIR maps and 333 MHz ($\lambda90$cm) maps respectively. Column 8  
are the data available at a higher frequency near 1 GHz ($\lambda20$cm): $^4$ 
\cite{braun07}, $^5$ VLA archival data using the CD array configuration 
(project code : AS325), $^6$ VLA archival map by combining interferometric data 
from C and D array, \cite{beck07}.   
\end{centering}
\label{tablesample} 
\end{table*}

So far, spatially resolved and global study of the correlation has been done
primarily using radio emission at 1.4 GHz and higher frequencies. The only low
frequency study done at 150 MHz \citep{cox88}, confirms that on global scales
the radio--FIR correlation holds good and is similar to what is seen at 1.4
GHz. To our knowledge, no low frequency ($<$ 1.4 GHz, such as 333 MHz)
spatially resolved study of the radio--FIR correlation exists in the
literature. The motivation to do such a study arises from the fact that at
lower frequencies the emission is largely nonthermal, hence better exhibiting
the relation between magnetic field and star formation. Secondly, since the
cosmic ray electrons (CRe) propagate larger distances in the galaxies at lower
frequencies, it is important to assess how that affects the form of the
radio--FIR correlation.

In this paper, we present spatially resolved study of the radio--FIR
correlation for four normal galaxies, NGC 4736, NGC 5055, NGC 5236 and NGC 6946
at spatial resolution of $\sim$1--1.5 kpc with radio observations made at 333
MHz ($\lambda90$ cm) and 1.4 GHz ($\lambda20$ cm). We also estimate the value
of $\kappa$ and verify the equipartition assumptions.  In Section~\ref{data} we
discuss the various sources of maps used in this work and also define the
parameter `$q$' which is used to quantify the correlation. In
Section~\ref{results} we present our results on spatially resolved radio--FIR
correlation using far infrared emission at $\lambda70~\mu$m and radio emission
at $\lambda20$ cm and $\lambda90$ cm. We discuss our results in
Section~\ref{discussion}.

\section{Data analysis}
\label{data}

The four galaxies in the sample for this study was chosen from \cite{basu12}.
The large angular size of the galaxies ensure enough independent regions to
carry out spatially resolved study. Our sample comprises of the galaxies NGC
4736, NGC 5055, NGC 5236 and NGC 6946. Table~\ref{tablesample} summarizes the
salient features of our sample and the various sources of obtaining the
archival data.

To study the radio--FIR correlation using nonthermal radio emission, a thorough
separation of thermal radio emission is needed.  We used nonthermal radio
continuum maps at $\lambda90$ cm and $\lambda20$ cm after separating the
thermal free--free component mainly originating from \hii regions in recent
star formation sites. Details of observation and data analysis are discussed in
\cite{basu12}. The thermal emission was estimated using the technique
developed by \cite{tabat07b}, wherein, the dust extinction-corrected H$\alpha$
map is used as a template for the thermal free-free emission. This is then
extrapolated to the desired radio frequency and subtracted from the total
emission map. The $\lambda90$ cm maps were obtained using the Giant Meterwave
Radio Telescope (GMRT) while the $\lambda20$ cm maps were obtained using
archival data from various assorted sources (see Table~\ref{tablesample}). The
nonthermal maps had 40 arcsec resolution with 9 arcsec pixel size.  We scaled
the flux of each galaxy to a common frequency of 1.4 GHz using the spectral
index map obtained from the 333 MHz and near 1 GHz images for each pixel.

The galaxies were observed in the far-infrared by the {\it Spitzer} at
$\lambda70~\mu$m as a part of the {\it Spitzer} Infrared Nearby Galaxy Survey
(SINGS; \citealt{kenni03}) using the Multiband Imaging Photometer for {\it
Spitzer} (MIPS; \citealt{rieke04}). The images were obtained from the publicly
available database in SINGS Data Release
5\footnote{http://data.spitzer.caltech.edu/popular/sings/}. The
$\lambda70~\mu$m images have a pixel size of 4.5 arcsec and a point spread
function (PSF) of about 16 arcsec. These were convolved to the resolution
of nonthermal radio maps (40 arcsec) and re-gridded to a common pixel size of
9 arcsec. All the maps were then aligned to the same coordinate system.

\begin{figure*}
\begin{center}
\begin{tabular}{cccc}
\multicolumn{4}{c}{{\bf \large Arm regions}}\\
{\mbox{\includegraphics[height=4cm, width=4cm]{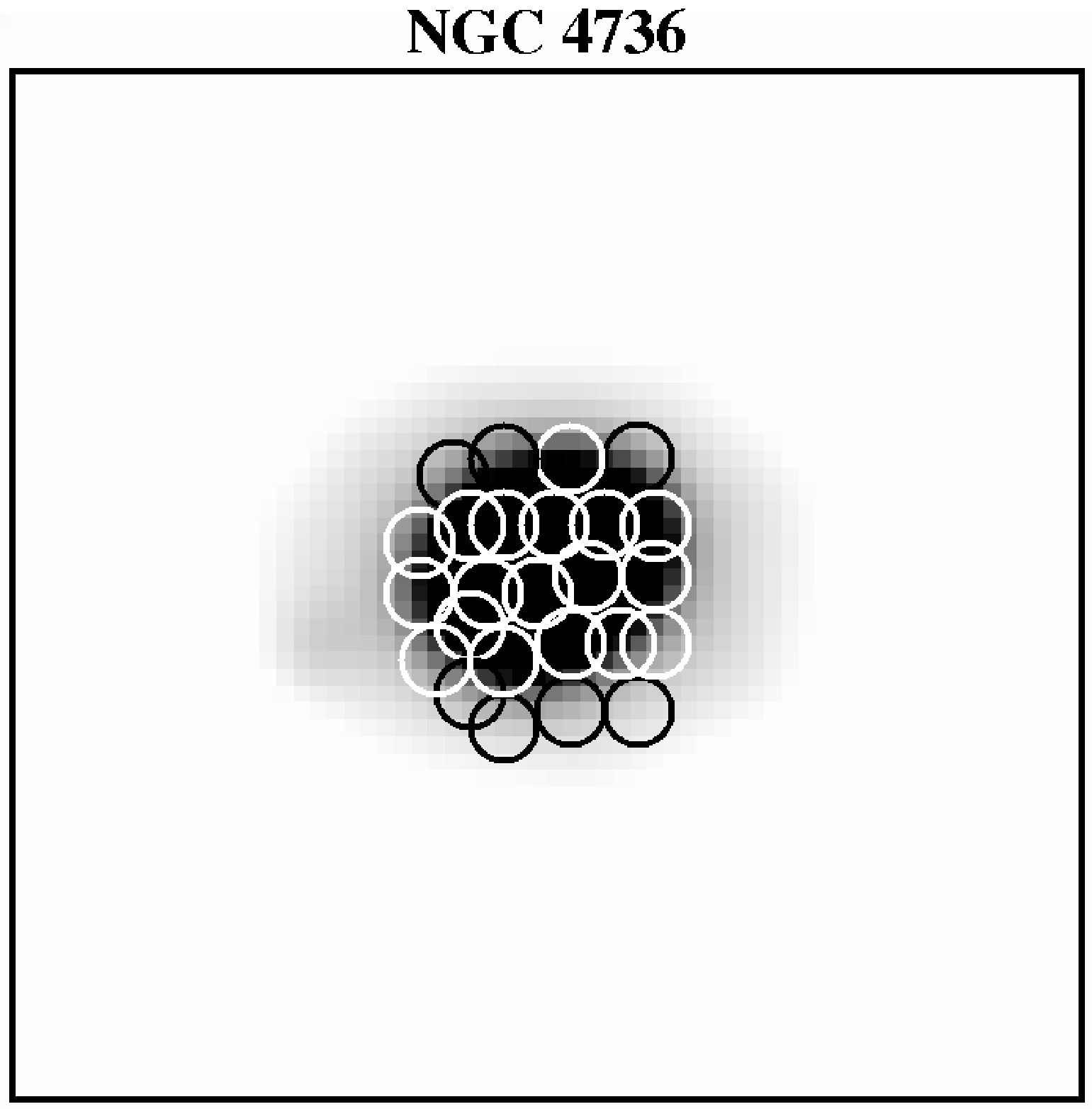}}} &
{\mbox{\includegraphics[height=4cm, width=4cm]{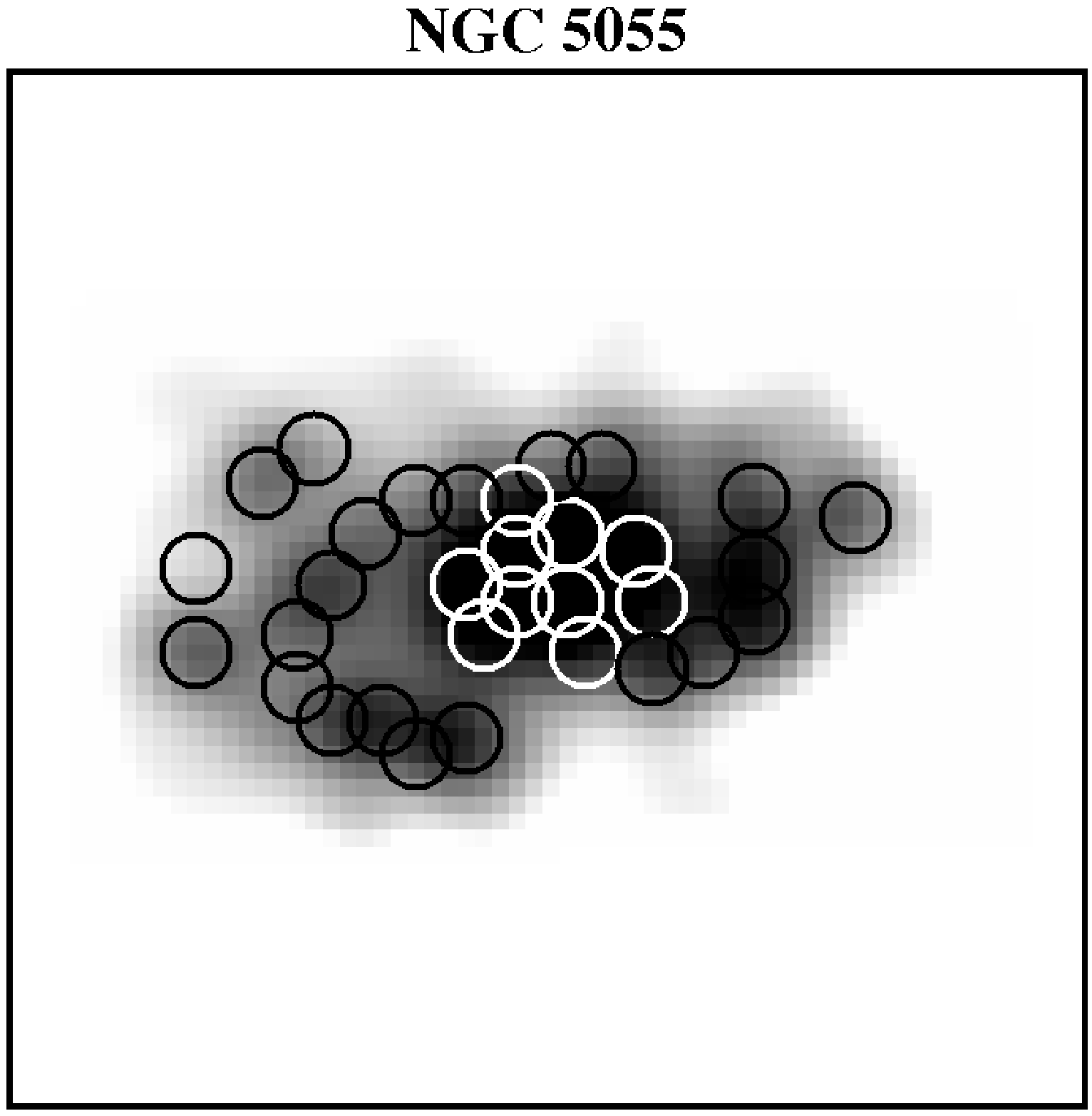}}} &
{\mbox{\includegraphics[height=4cm, width=4cm]{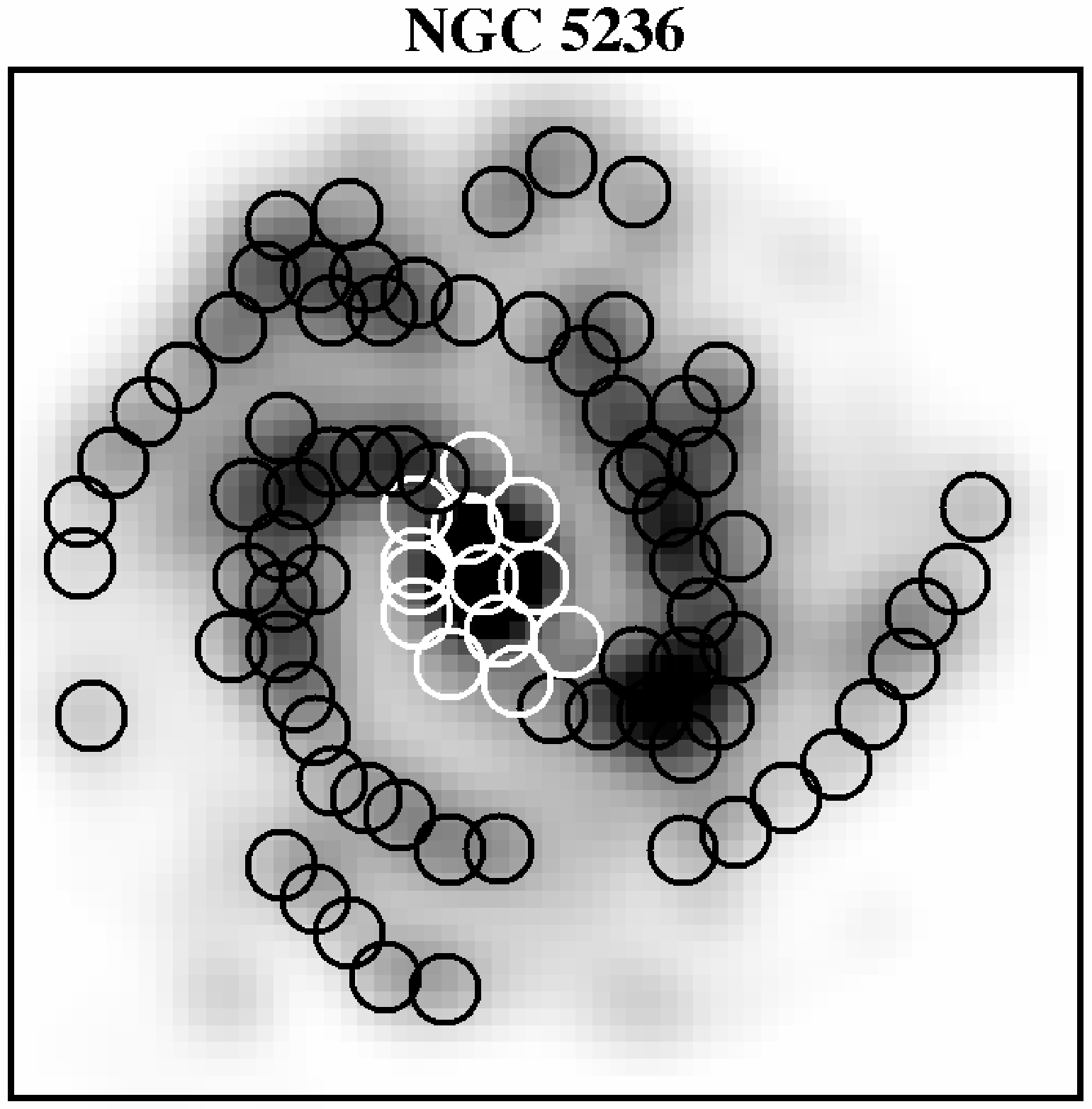}}} &
{\mbox{\includegraphics[height=4cm, width=4cm]{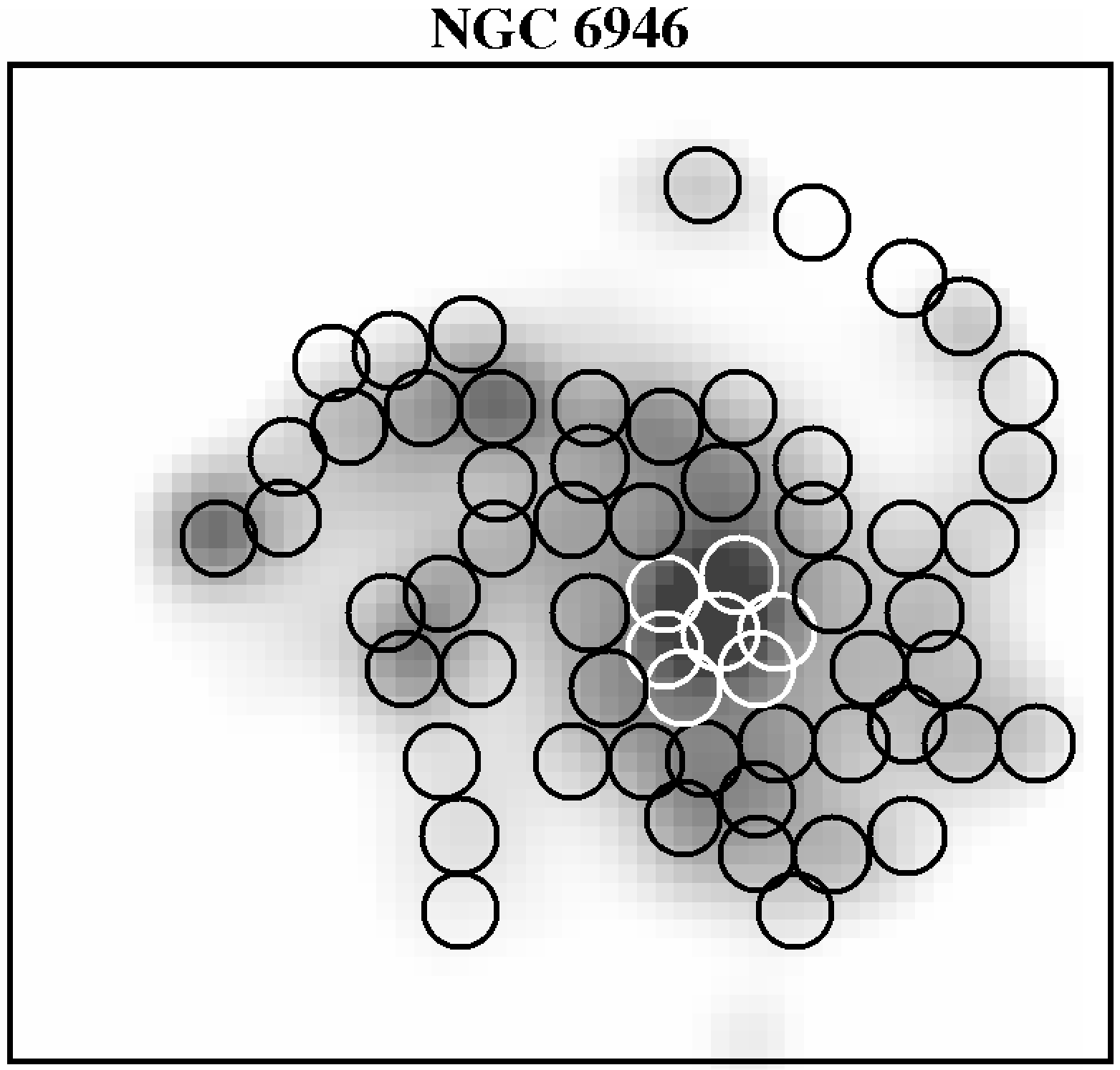}}} \\
&&&\\
\multicolumn{4}{c}{{\bf \large Interarm regions}}\\
{\mbox{\includegraphics[height=4cm, width=4cm]{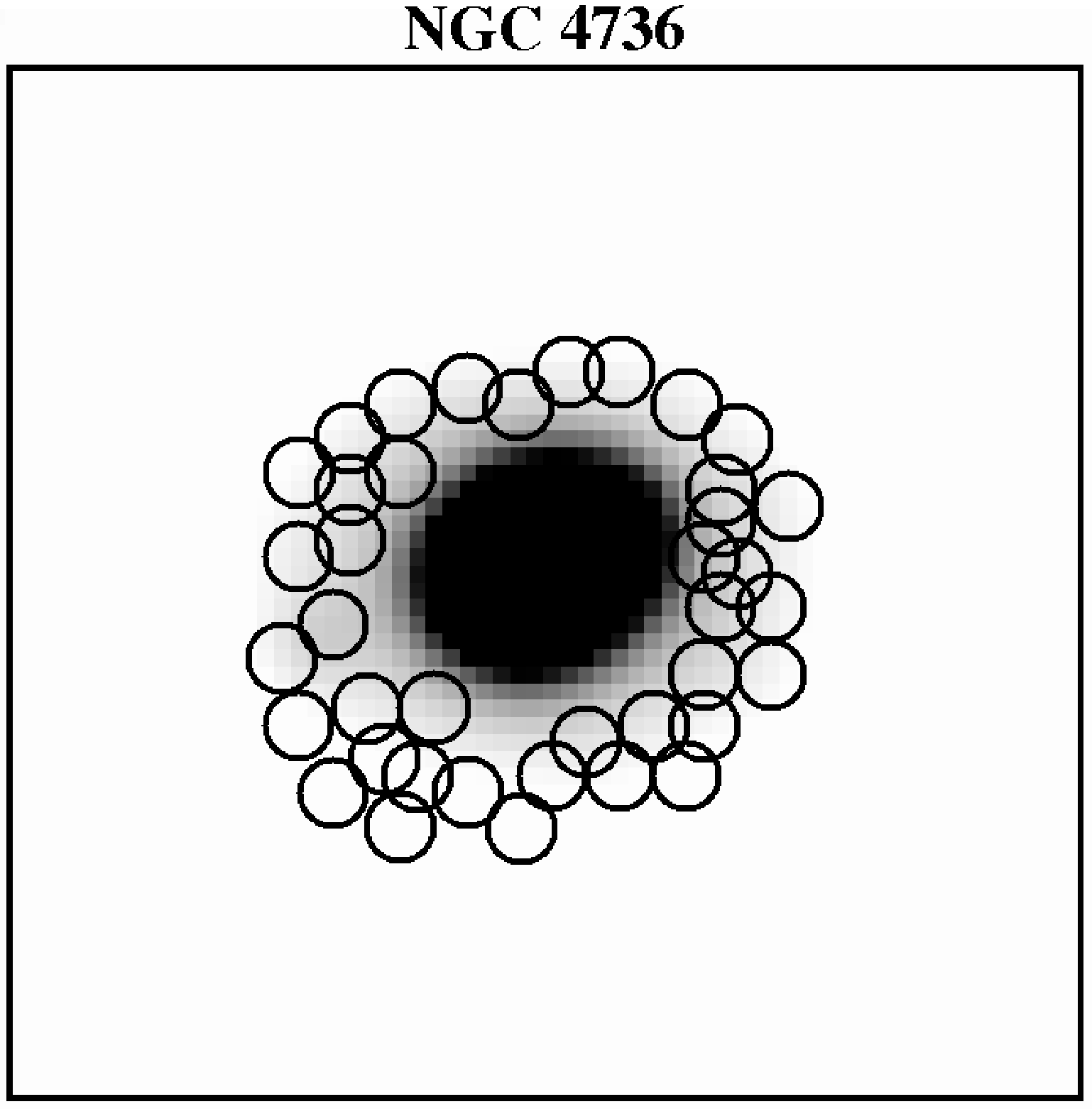}}}&
{\mbox{\includegraphics[height=4cm, width=4cm]{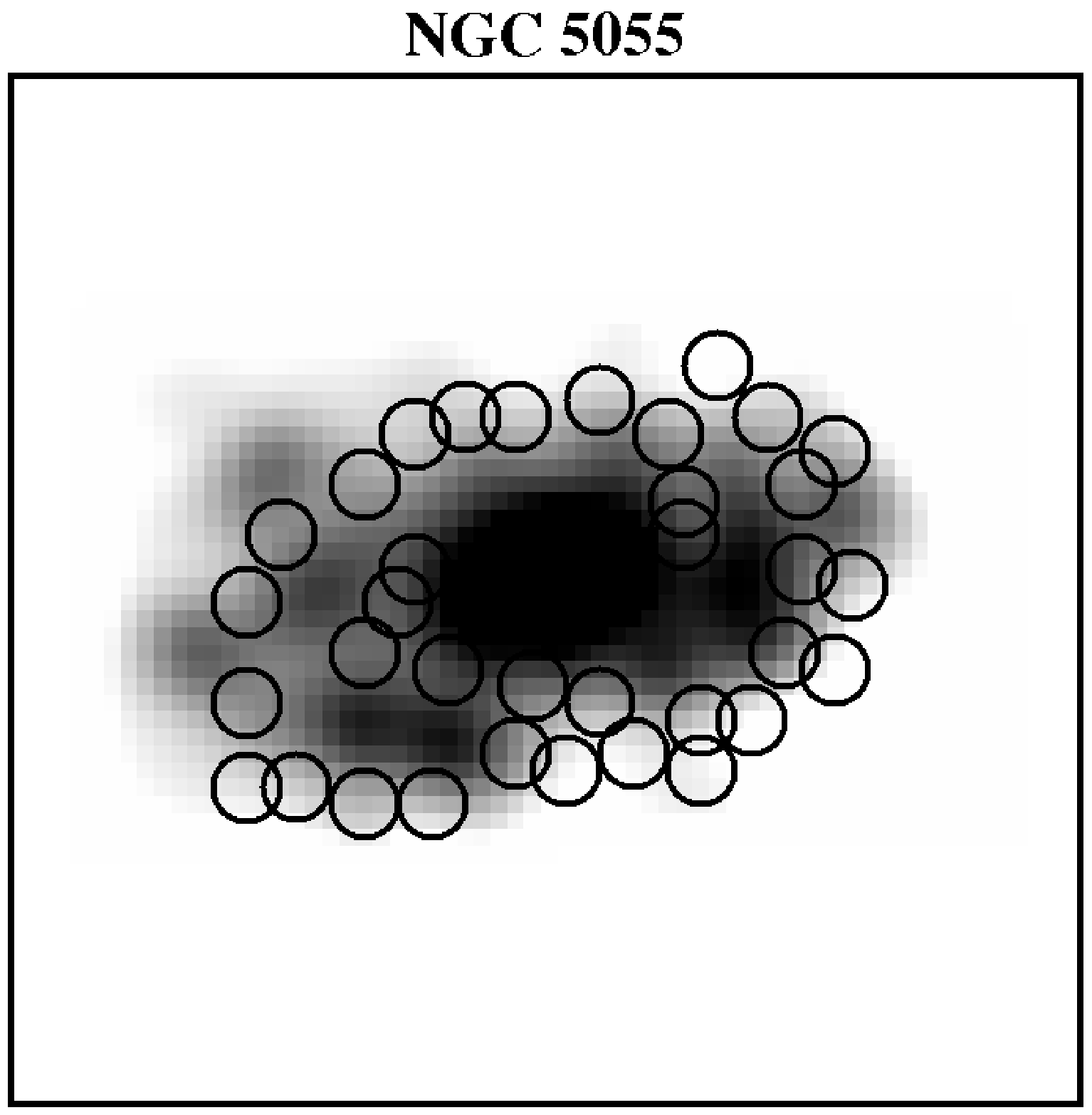}}}&
{\mbox{\includegraphics[height=4cm, width=4cm]{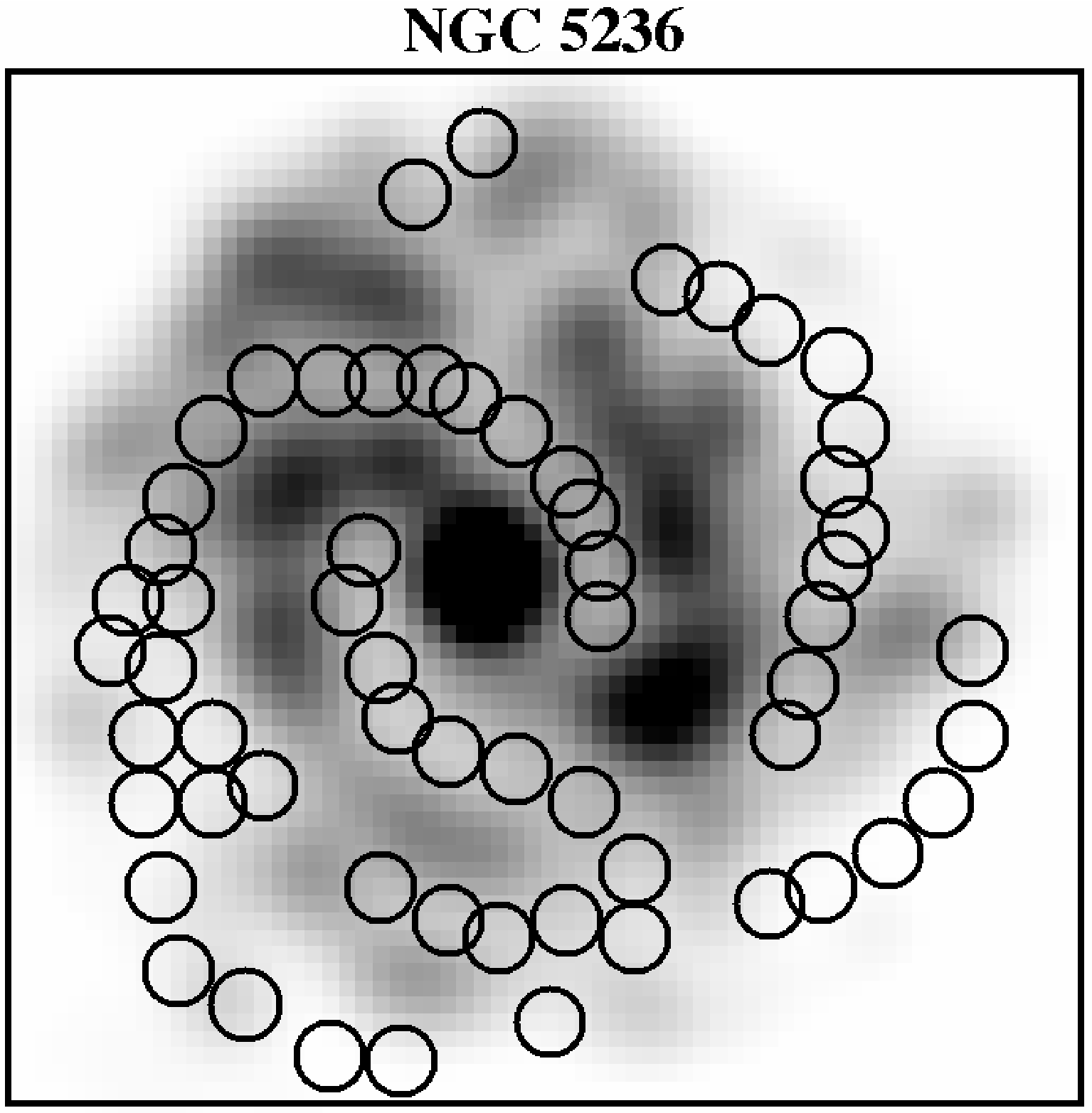}}}&
{\mbox{\includegraphics[height=4cm, width=4cm]{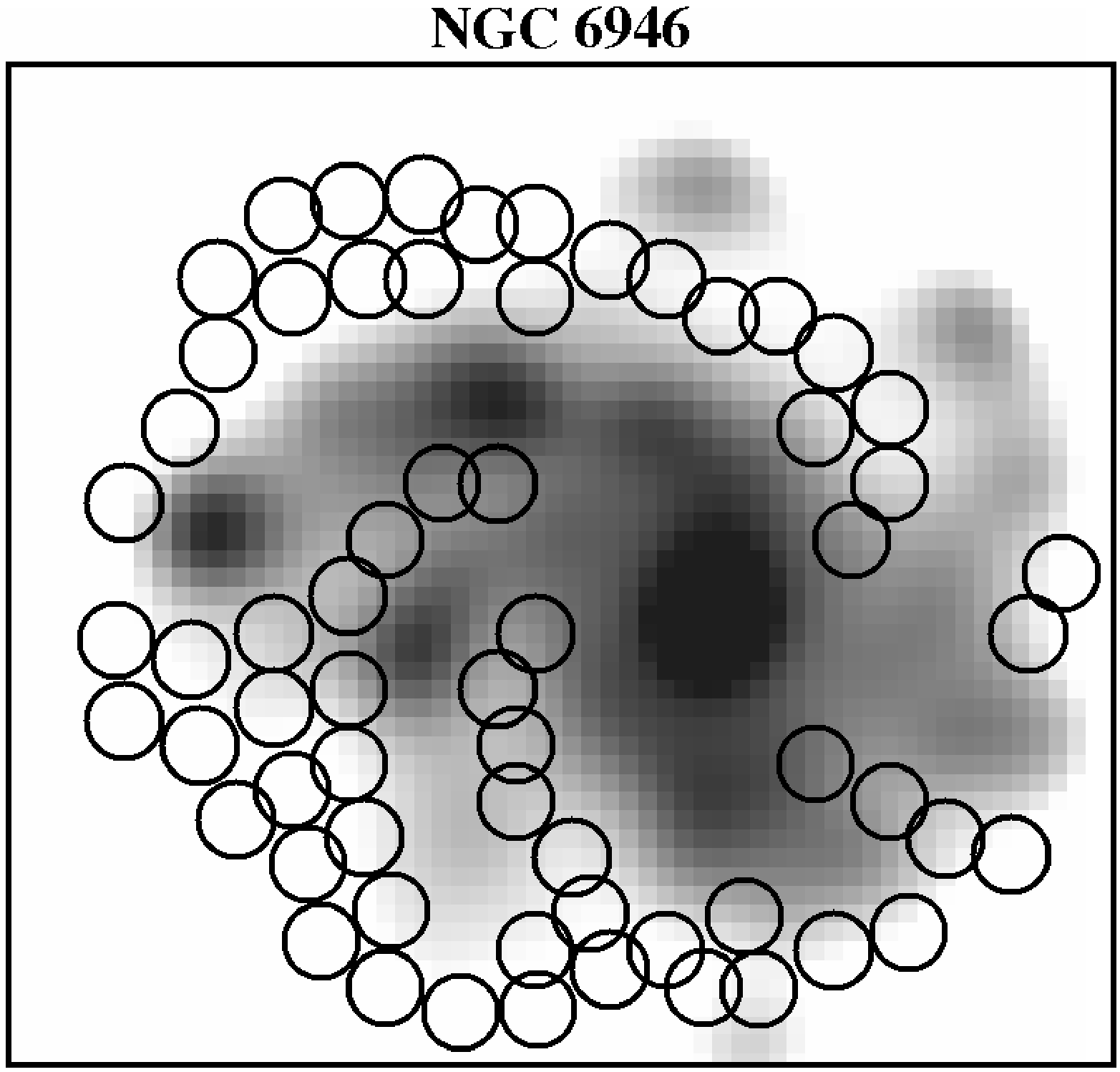}}} \\
\end{tabular}
\caption{Overlay of $\sim$40 arcsec beams (marked in circles) on the H$\alpha$
images$^a$ smoothed to 40 arcsec. The top and lower panels shows the arm and
the interarm regions respectively (see Section~3 for details).\\
$^a$The images were downloaded from the NED for the galaxies
NGC 4736 \citep[1-m Jacobus Kapteyn Telescope (JKT) at La Palma with filter
Ha6570;][]{knape04}, NGC 5055 (2.3-m telescope at KPNO, filter: 6580) and NGC
5236 (0.9-m telescope at CTIO, filter: 6563). For NGC 6946 the image was
downloaded from the SINGS website.}
\label{region} \end{center}
\end{figure*}

\begin{table*}
\caption{Integrated flux densities of the galaxies at $\lambda$90 cm (333 MHz;
\citealt{basu12}), $\lambda$20 cm (1400 MHz; spectral index scaled from data
given in column 8 of Table~\ref{tablesample}) and $\lambda70~\mu$m (taken
from the NED). The map noise ($\sigma$)for the 40 arcsec resolution images are
also given.}
\begin{tabular}{@{}ccccccccc@{}}
 \hline
Name& $S_{\rm 90cm}$  & $\sigma_{\rm 90cm}$ & $S_{\rm 20cm}$ & $\sigma_{\rm 20cm}$ & $S_{\rm 70\mu m}$& $\sigma_{\rm 70\mu m}$\\
& Jy& mJy beam$^{-1}$&Jy&mJy beam$^{-1}$&Jy&mJy beam$^{-1}$\\
\hline 
NGC 4736&0.9$\pm$0.06& 2 &0.31$\pm$ 0.03&0.35 &93.93$\pm$ 7.34&15\\
NGC 5055&2.3$\pm$0.13&3& 0.41$\pm$ 0.05&0.4&72.57$\pm$ 5.16&15\\
NGC 5236&6.86$\pm$0.62&2.5& 2.36$\pm$ 0.18&0.3&312.0$\pm$ 15.6&30\\
NGC 6946&4.3$\pm$0.24&1&1.5$\pm$ 0.1&0.2&207.2$\pm$ 16.1&25\\
\hline 
\end{tabular}
\label{tableflux}
\end{table*}

For the present study, the flux density per beam for the radio and FIR
maps were determined within an area of $\sim40$ arcsec diameter,
with the adjacent region being about one beam away to ensure independence.
Pixels with brightness above 2$\sigma$ ($\sigma$ is the $rms$ noise in the map)
were considered for the analysis. We estimate the slope of the radio--FIR and
the quantity `$q$' introduced by \cite{helou85}.

The parameter $q$ is used as a measure of the radio--FIR correlation, where its
dispersion indicates the tightness of the correlation. Conventionally it is
defined as the logarithm of the ratio of total FIR flux between
$\lambda40~\mu$m and $\lambda120~\mu$m and the radio flux measured at 1.4 GHz.
However, we define $q$ as per \citet[][]{apple04} using FIR flux density at
$\lambda70~\mu$m, such that, 
$$
q_{\lambda} = \log_{10} \left( S_{\rm 70\mu m}/S_{\lambda} \right)
$$
where, $\lambda$ is the radio wavelength (here, $\lambda= $ 20 cm or 90 cm) and
$S_{\rm 70\mu m}$ and $S_{\lambda}$ are the flux densities of $\lambda70~\mu$m
and radio wavelength respectively. The FIR emission from galaxies between
$\lambda40~\mu$m and $\lambda120~\mu$m is dominated by the emission from cool
dust with dust temperature, $T_{\rm dust}\sim$20 K \citep[see e.g,][]{xu92,
hoern98, tabat07b, basu12}. The peak of this emission occurs at about
$\lambda100~\mu$m.  Note that a black body at $\sim$20 K peaks at about
$\lambda145~\mu$m, but a grey body ($\lambda^{-\beta}B_\lambda(T)$, where
$\beta=2$ is the dust emissivity index and $B_\lambda(T)$ is the Planck
function) has a peak at $\sim\lambda100~\mu$m. The maps at $\lambda70~\mu$m,
which are nearest available to $\lambda100~\mu$m, essentially traces this
component of the dust. Using monochromatic $\lambda70~\mu$m emission to study
the radio--FIR correlation does not affect the conclusions significantly,
except for a slight increase in the dispersion \citep{murph06a}.

\begin{figure*}
\begin{center}
\begin{tabular}{cc}
{\mbox{\includegraphics[scale=0.26]{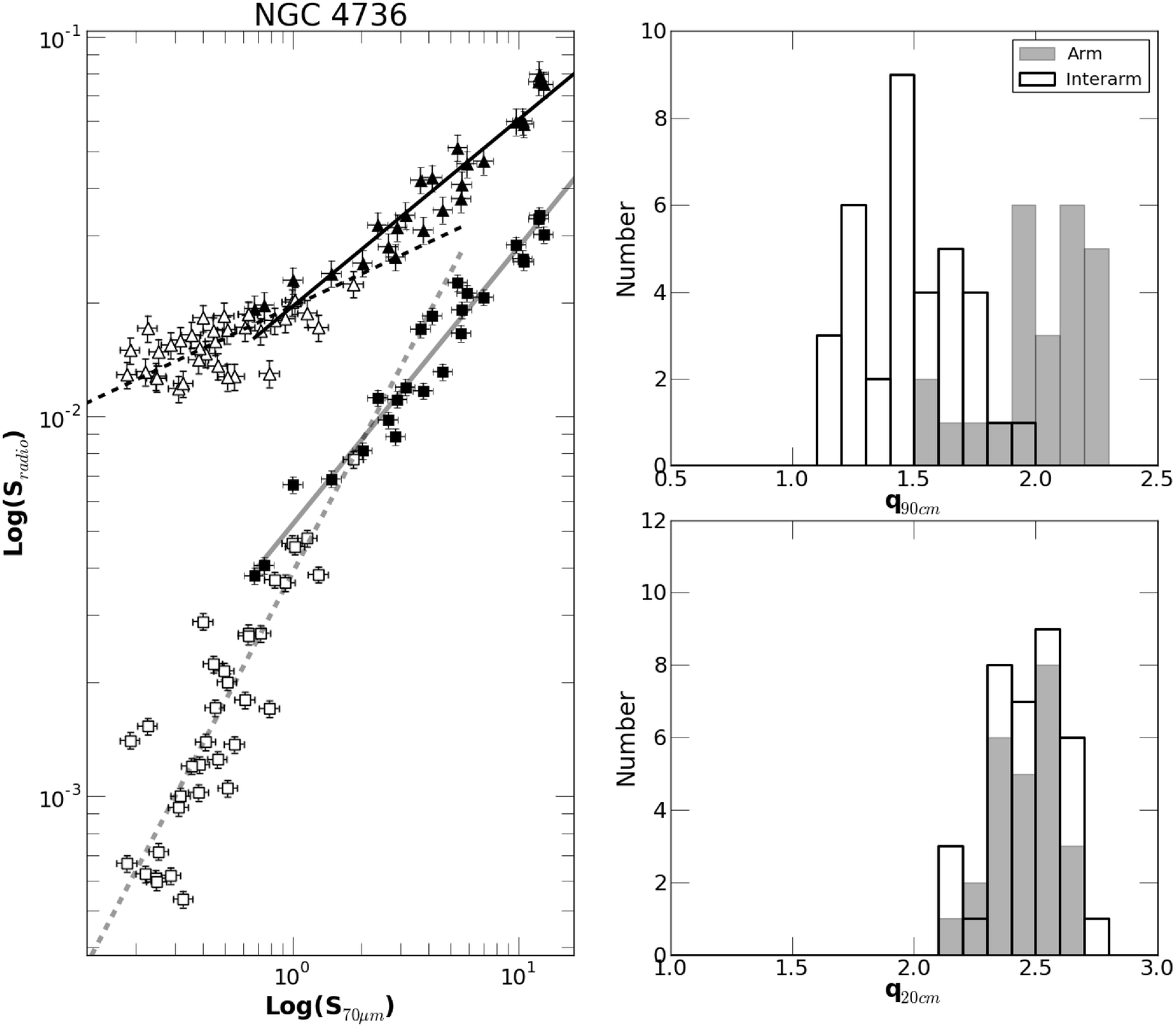}}} &
{\mbox{\includegraphics[scale=0.26]{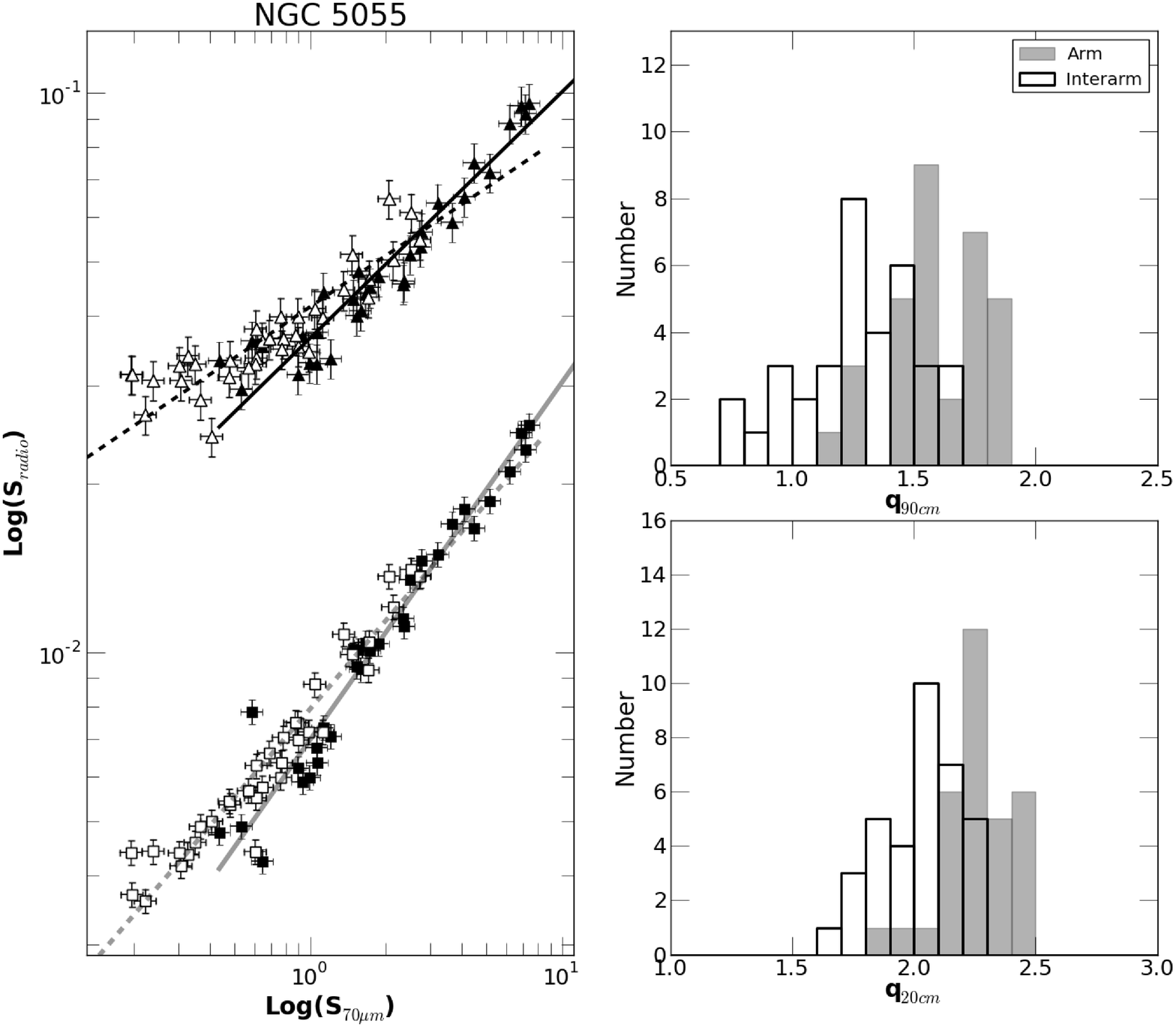}}} \\
{\mbox{\includegraphics[scale=0.26]{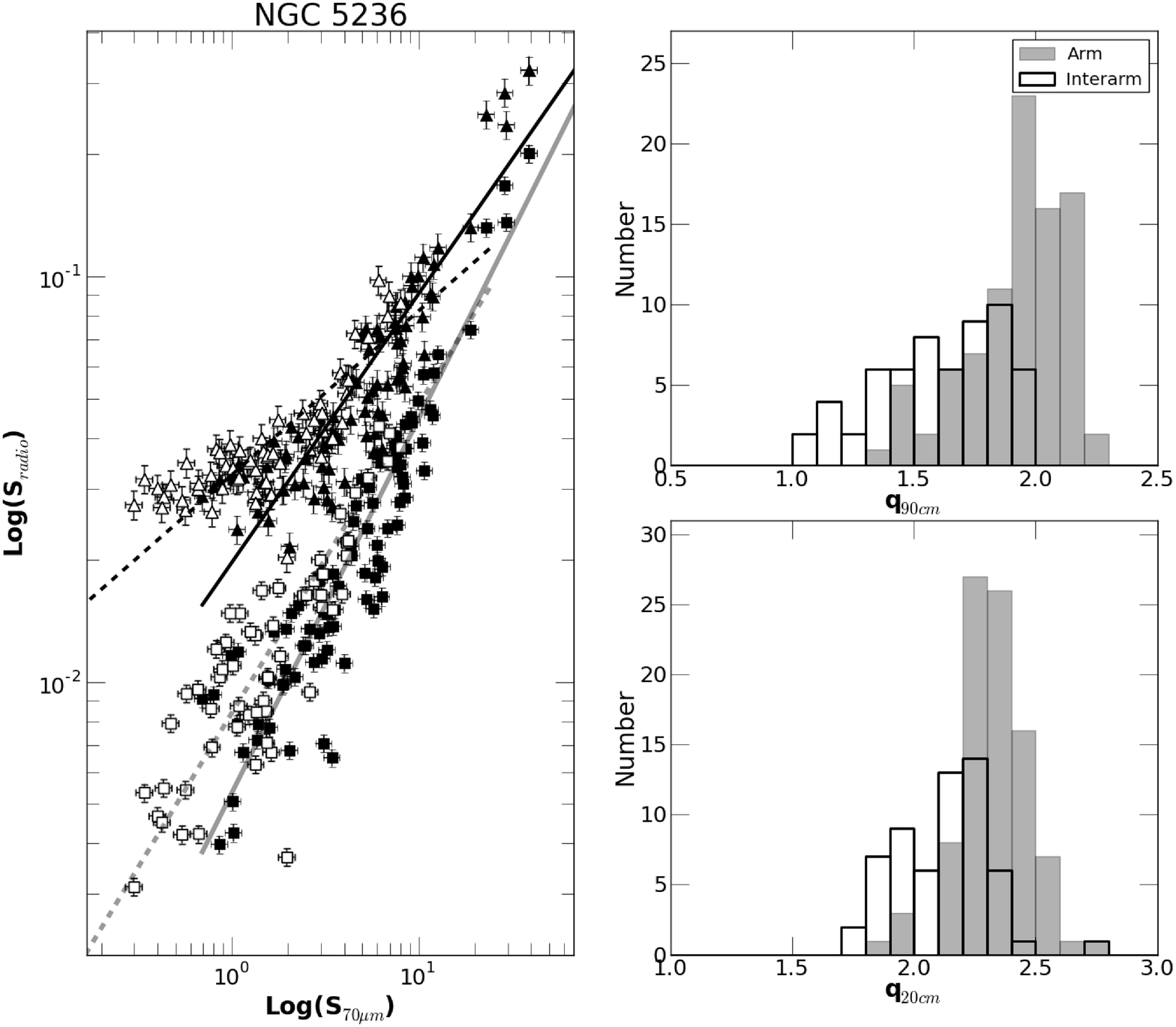}}} &
{\mbox{\includegraphics[scale=0.26]{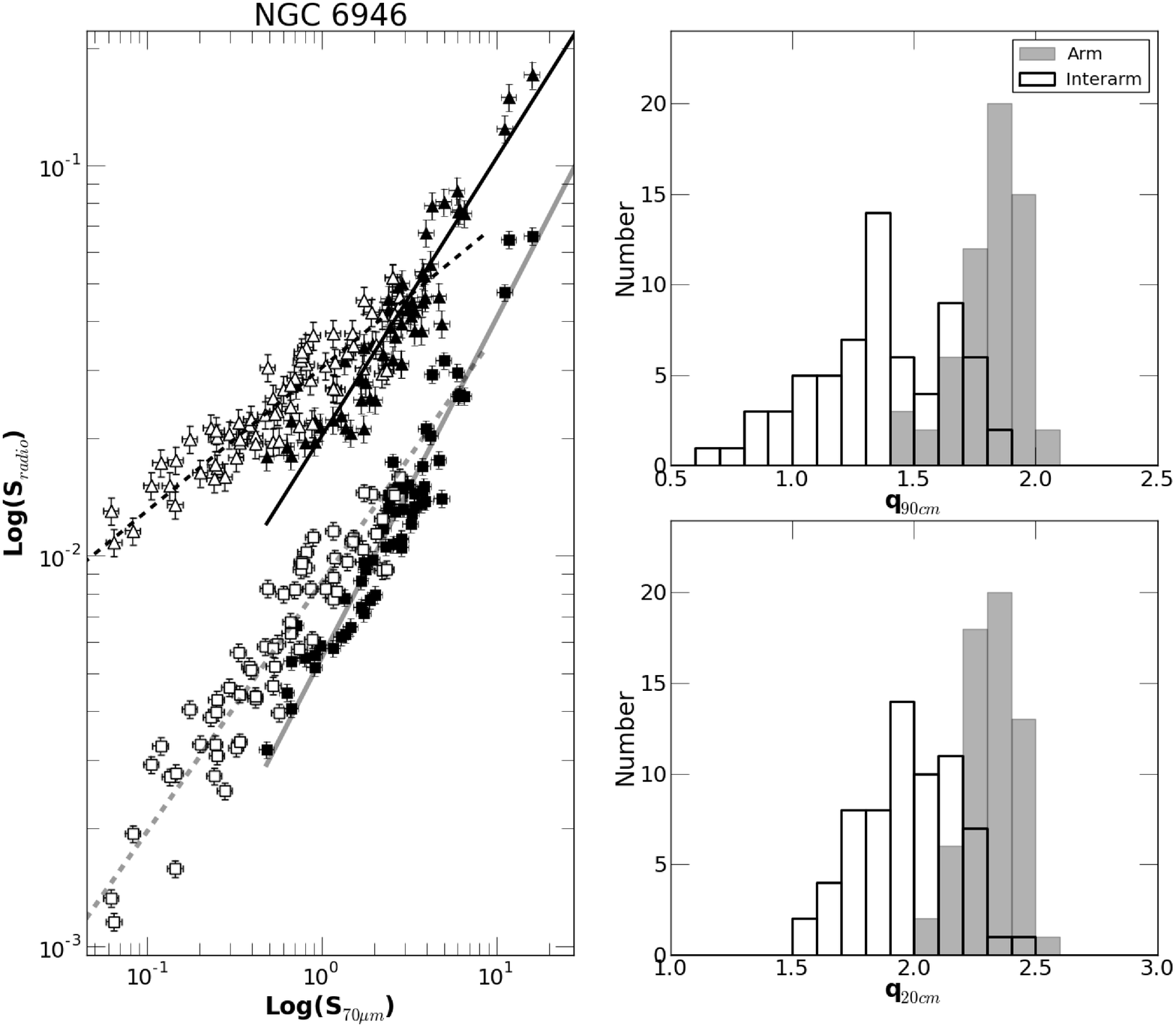}}} \\
\end{tabular}
\caption{The figure shows the radio intensity vs. $\lambda70~\mu$m FIR
intensity (in Jy beam$^{-1}$). The triangles are for $\lambda90$ cm and
squares are for $\lambda20$ cm.  The filled symbols are for arms and unfilled
symbols are for interarms.  The histograms are the distribution of $\qpband$
and $\qlband$, where arms are shown with filled grey and interarms with
unfilled histograms. The lines are the fit to the data of the form $S_{\rm
radio} = a\times S_{\rm 70\mu m}^b$ (See Table 3). The solid and dashed lines
are fit to the arm and interarm regions. Black lines are for $\lambda90$ cm,
while grey lines are for $\lambda20$ cm.}
\label{radio-fir}
\end{center}
\end{figure*}

\begin{table*}
\begin{centering}
\caption{Summary of the values of $q_\lambda$ and the fitted parameters for
each of the galaxies. Columns 3, 4 and 5, 6 are the mean values of $q_\lambda$
and their dispersion at $\lambda20$cm and $\lambda90$cm respectively as shown
in Figure~\ref{radio-fir}. The value of $q_\lambda$ was computed using the flux
density within one beam of FWHM $\sim40$ arcsec.  Columns 7 and 9 are the
fitted values for $q_\lambda$, while Columns 8 and 10 are the slopes of the
radio--FIR correlation. Here, $a_{20\rm cm}$, $b_{20\rm cm}$ and $a_{90\rm
cm}$, $b_{90\rm cm}$ are the parameters $a$ and $b$ in the fitted equation,
$S_{\rm radio} = aS_{\rm IR}^b$, at wavelengths 20 cm and 90 cm respectively.}
\begin{tabular}{@{}llccccccccccc@{}}
 \hline
 \multicolumn{1}{l}{Name}  &
 \multicolumn{1}{c}{}  &
 \multicolumn{1}{c}{$\langle q_{20\rm cm}\rangle$}  &
 \multicolumn{1}{c}{$\sigma_{q_{20\rm cm}}$}  &
 \multicolumn{1}{c}{} &
 \multicolumn{1}{c}{} &
 \multicolumn{1}{c}{$\langle q_{90\rm cm}\rangle$} &
 \multicolumn{1}{c}{$\sigma_{q_{90\rm cm}}$}&
 \multicolumn{1}{c}{} &
 \multicolumn{4}{c}{$S_{\rm radio} = a S_{\rm IR}^b$}\\
\cline{10-13}
&&&&&&&&& $-\log(a_{20\rm cm})$ &  $b_{20\rm cm}$ & $-\log(a_{90\rm cm})$ &  $b_{90\rm cm}$ \\
~~~~(1)&(2)&(3)&(4)&&&(5)&(6)&&(7)&(8)&(9)&(10) \\

\hline
NGC 4736	& arm:		& 2.45	& 0.12	&	&	& 2.01	& 0.20 &	& 2.28$\pm$0.02	& 0.73$\pm$0.04	& 1.71$\pm$0.03	& 0.49$\pm$0.04	\\
		& interarm:	& 2.46	& 0.15	&	&	& 1.46	& 0.23 &	& 2.41$\pm$0.04	& 1.13$\pm$0.12	& 1.71$\pm$0.03	& 0.28$\pm$0.04	\\
&&&&&&&&&&&&\\
NGC 5055	& arm:		& 2.26	& 0.14	&	&	& 1.59	& 0.20 &	& 2.14$\pm$0.02	& 0.64$\pm$0.03	& 1.44$\pm$0.02	& 0.44$\pm$0.03	\\
		& interarm:	& 2.03	& 0.16	&	&	& 1.27	& 0.24 &	& 2.10$\pm$0.01	& 0.53$\pm$0.03	& 1.38$\pm$0.01	& 0.30$\pm$0.04	\\
&&&&&&&&&&&&\\
NGC 5236 	& arm:		& 2.32	& 0.14	&	&	& 1.92	& 0.20 &	& 2.27$\pm$0.04	& 0.92$\pm$0.04	& 1.71$\pm$0.03	& 0.66$\pm$0.04	\\
	 	& interarm:	& 2.13	& 0.20	&	&	& 1.60	& 0.26 &	& 2.08$\pm$0.03	& 0.76$\pm$0.05	& 1.50$\pm$0.02	& 0.40$\pm$0.04	\\
&&&&&&&&&&&&\\
NGC 6946	& arm:		& 2.31	& 0.10	&	&	& 1.81	& 0.14 &	& 2.26$\pm$0.02	& 0.87$\pm$0.04	& 1.69$\pm$0.02	& 0.71$\pm$0.04	\\
		& interarm:	& 1.97	& 0.19	&	&	& 1.32	& 0.32 &	& 2.07$\pm$0.01	& 0.64$\pm$0.03	& 1.52$\pm$0.01	& 0.34$\pm$0.02	\\
\hline 
\end{tabular}
\end{centering}

\label{tableqlambda}
\end{table*}

\section{Results}
\label{results}

The spatially resolved study of the radio--FIR correlation was done by
broadly classifying the emission from arm (including the central region) and
interarm regions of these galaxies. The arms were identified from the
H$\alpha$ images for each galaxy.  For the ringed galaxy NGC 4736, which
has no prominent arms, the star forming ring was taken as the arm. The arm and
the interarm regions used in our analysis are plotted as circles which are
overlayed on the 40 arcsec H$\alpha$ images in Figure~\ref{region}.  The
quantity $q_\lambda$ was computed within each such region. Note that the
calibration uncertainty at $\lambda70~\mu$m could be $\sim20$ per cent
\citep{murph06a}.  This would lead to a systematic error of about 10 per cent
in the values of $q_\lambda$.

Table 2 gives the total flux density (in Jy) and map rms noise of the 40
arcsec images (in mJy beam$^{-1}$) of the galaxies.  The galaxy integrated
mean values of $\qlband$ are $2.48\pm0.1$, $2.25\pm0.07$, $2.12\pm0.06$ and
$2.14\pm0.07$, for the galaxies NGC 4736, NGC 5055, NGC 5236 and NGC 6946
respectively. The $\qpband$ are $2.02\pm0.07$, $1.5\pm0.1$, $1.66\pm0.09$ and
$1.68\pm0.08$ respectively.  However, the spatially resolved estimates of
$\qlband$ and $\qpband$ suggests that their values vary between arm and
interarm regions. Figure~\ref{radio-fir} shows the brightness of the
nonthermal radio emission with the far infrared emission at $\lambda70~\mu$m,
both in units of Jy beam$^{-1}$, for all the four galaxies.  The figure also
shows the distribution of $q_\lambda$ for $\lambda20$ cm and $\lambda90$ cm.
It was seen that the star forming, gas rich spiral arms of the galaxies
showed higher values for $q_\lambda$ when compared to the adjacent low star
forming interarm regions. Table 3 summarizes the mean value of the quantity
$q_{\lambda}$ and its dispersion for arm and interarm regions.

\begin{figure}
\begin{center}
\begin{tabular}{c}
\includegraphics[width=8cm,height=5.5cm]{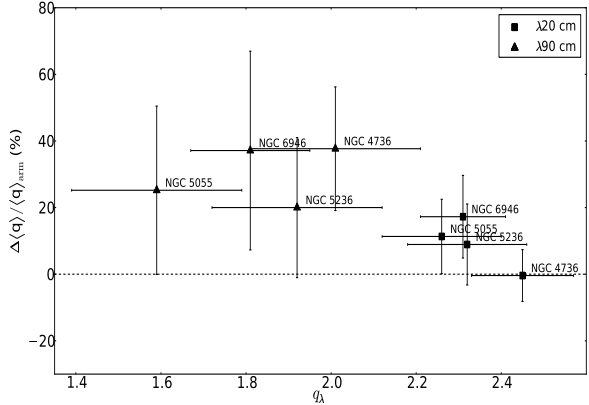}
\end{tabular}
\caption{The percentage change in the value of $\langle q_{\lambda}\rangle$
between arms and interarms.  Here, $\Delta \langle q\rangle = \langle
q_{\lambda}\rangle_{\rm arm} - \langle q_{\lambda}\rangle_{\rm interarm}$ for
each of the sample galaxies. The triangles are for $\lambda90$cm and squares
are for $\lambda20$cm. The mean change in the value of $\qlband$ between arms
and interarms is $\sim 10$ percent, while $\qpband$ changes by $\sim30$
percent.}
\label{qdiff}
\end{center}
\end{figure}

The mean value in the arms for all the galaxies were found to be, $\langle
\qlband\rangle_{\rm arm} = 2.32$  with a narrow dispersion of $\sigma_{\qlband,
\rm arm} = 0.14$, while for the interarms $\langle \qlband\rangle_{\rm
interarm} = 2.15$ and $\sigma_{\qlband, \rm interarm} = 0.3$. At $\lambda90$cm,
we find the $\langle \qpband\rangle_{\rm arm} = 1.85$ with $\sigma_{\qpband,
\rm arm} = 0.22$ and $\langle \qpband\rangle_{\rm interarm} = 1.43$ and
$\sigma_{\qpband, \rm interarm} = 0.3$. Figure~\ref{qdiff} shows the percentage
change in the value of $q_\lambda$ between arms and interarms for each of the
galaxies, where $\Delta q = \langle q_\lambda\rangle_{\rm arm} - \langle
q_\lambda\rangle_{\rm interarm}$. The squares are for $\lambda20$ cm and
triangles are for $\lambda90$ cm.  The mean of the $\qlband$ changes slightly,
by about 9 percent, between arms and interarms, however, this change is
significant with $>$ 99.9 percent confidence using Kolmogorov-Smirnov test.  At
$\lambda90$ cm, the change in the value of $\langle q_\lambda\rangle$ between
arm and interarm regions are much pronounced with $\Delta q\sim$ 30
percent.

\subsection{Fit to the radio and IR flux densities}

The data were fitted using the form $S_{\rm radio} = a\times S_{\rm IR}^b$,
where, $S_{\rm radio}$ is the flux density of the radio emission at
$\lambda20$cm and $\lambda90$cm, and $S_{\rm IR}$ is the flux density of the
$\lambda70\mu$m infrared emission. The fitting parameter `$a$' is an estimate
for $q_{\lambda}$, where $q_{\lambda} = -\log_{10}a$, such that $\log_{10}
S_\lambda = -q_\lambda + b\times\log_{10}S_{IR}$. The slope of the radio-infrared
correlation is given by the parameter `$b$'. Separate fits were done for arm
and interarm regions using ordinary least-square `bisector method'
\citep{isobe90} in the log-log plane. The parameters obtained are summarized in
Table 2 in the last four columns for $\lambda20$ cm and $\lambda90$ cm. The
values of $\qlband$ and $\qpband$ are in good agreement with the mean values
obtained from the respective distribution. The fitted parameters are plotted in
Figure~\ref{radio-fir}.  The black lines are for fits at $\lambda90$cm and the
grey lines are at $\lambda20$cm. The solid lines are fit to the arm regions
only and the dashed lines are for the interarms.  All the correlations are
highly significant in our case with Pearson's correlation coefficient, $r >
0.8$ (and $r > 0.9$ in most of the cases), except for interarm regions of NGC
4736 at $\lambda90$ cm, where $r = 0.68$.

The slope of the $\lambda20$ cm and $\lambda70~\mu$m flux density for the arm
regions for all the galaxies lies between $\sim 0.65$ -- 0.9. However, for the
interarm regions the slope is slightly shallower, lying in the range 0.55 -- 1.
The mean value of the parameters for arm and interarm after the fit can be
written as,
\begin{eqnarray}
&& \log_{10} S_{20\rm cm} = {-(2.24\pm0.05)} + (0.8\pm0.08)\log_{10} S_{70\mu\rm m} \nonumber\\
&& \quad\quad\quad\quad\quad\quad\quad\quad\quad\quad\quad\quad\quad\quad\quad\quad{\rm for~arm} 
\label{fit20arm}
\end{eqnarray}
\begin{eqnarray}
&&  \log_{10} S_{20\rm cm} = {-(2.17\pm0.05)} + (0.76\pm0.14)\log_{10} S_{70\mu\rm m} \nonumber\\
&& \quad\quad\quad\quad\quad\quad\quad\quad\quad\quad\quad\quad\quad\quad\quad\quad{\rm for~interarm} 
\label{fit20iarm}
\end{eqnarray}
The fitted value of $\qlband$ and the slope differs slightly from arms
to interarms.

At $\lambda90$cm, we find that the slope lies in the range $\sim$ 0.45 -- 0.7
for the arms, whereas in the interarm region the slope lies in the range $\sim$
0.3 -- 0.4. The slopes are much flatter than at $\lambda20$cm. The mean values
of the fitted parameters are found to be,
\begin{eqnarray}
&& \log_{10} S_{90\rm cm} = {-(1.64\pm0.05)} + (0.6\pm0.1)\log_{10} S_{70\mu\rm m} \nonumber\\
&& \quad\quad\quad\quad\quad\quad\quad\quad\quad\quad\quad\quad\quad\quad\quad\quad{\rm for~arm} 
\label{fit90arm}
\end{eqnarray}
\begin{eqnarray}
&& \log_{10} S_{90\rm cm} = {-(1.53\pm0.04)} + (0.33\pm0.07)\log_{10} S_{70\mu\rm m}  \nonumber\\
&& \quad\quad\quad\quad\quad\quad\quad\quad\quad\quad\quad\quad\quad\quad\quad\quad {\rm for~interarm}
\label{fit90iarm}
\end{eqnarray}
There is a significant change in the value of $\qpband$ and the slope between the arm 
and the interarm regions. 

\subsection{q vs. $\ant$}

\begin{figure*}
\begin{center}
\begin{tabular}{cc}
{\mbox{\includegraphics[scale=0.25]{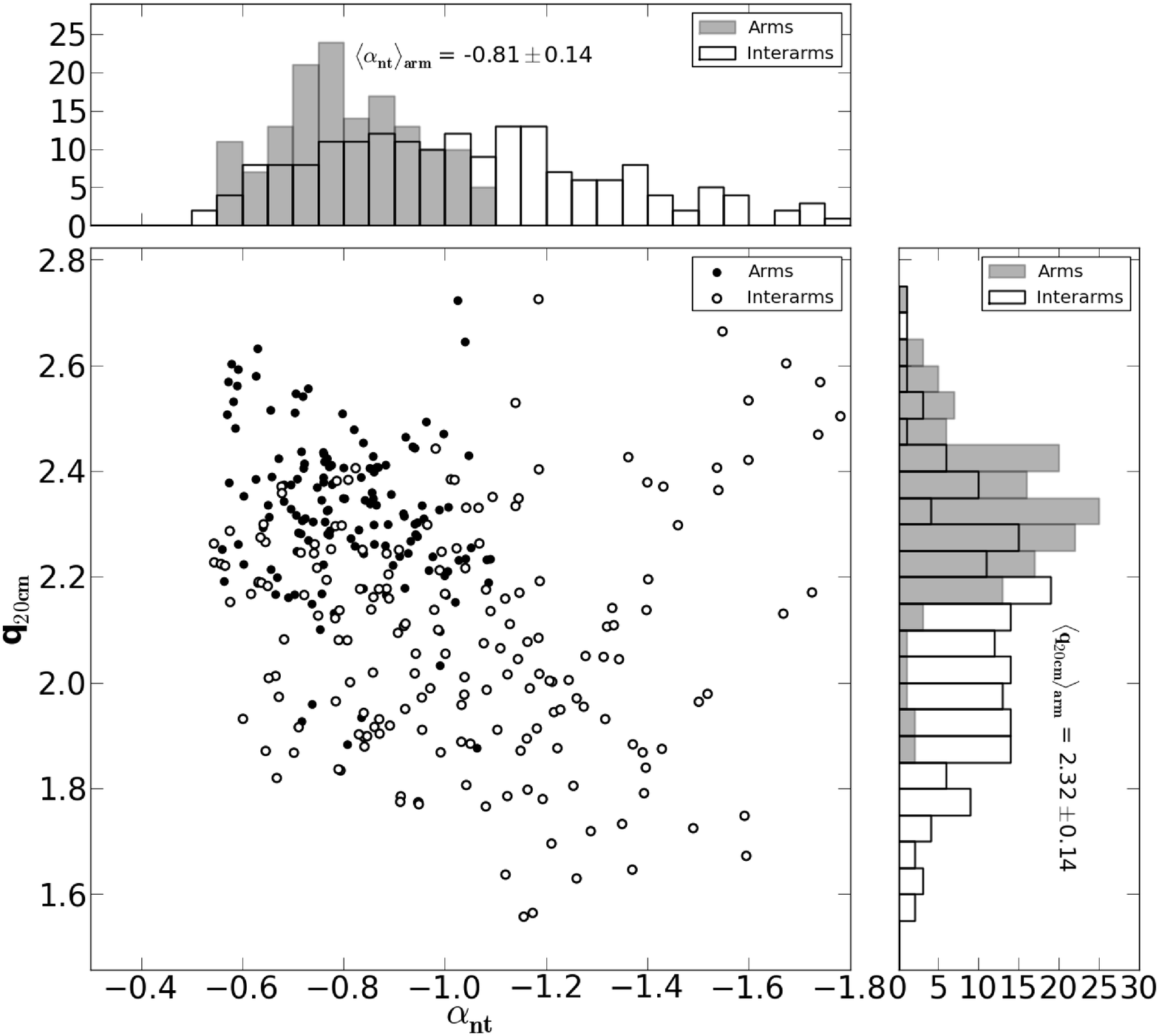}}} &
{\mbox{\includegraphics[scale=0.25]{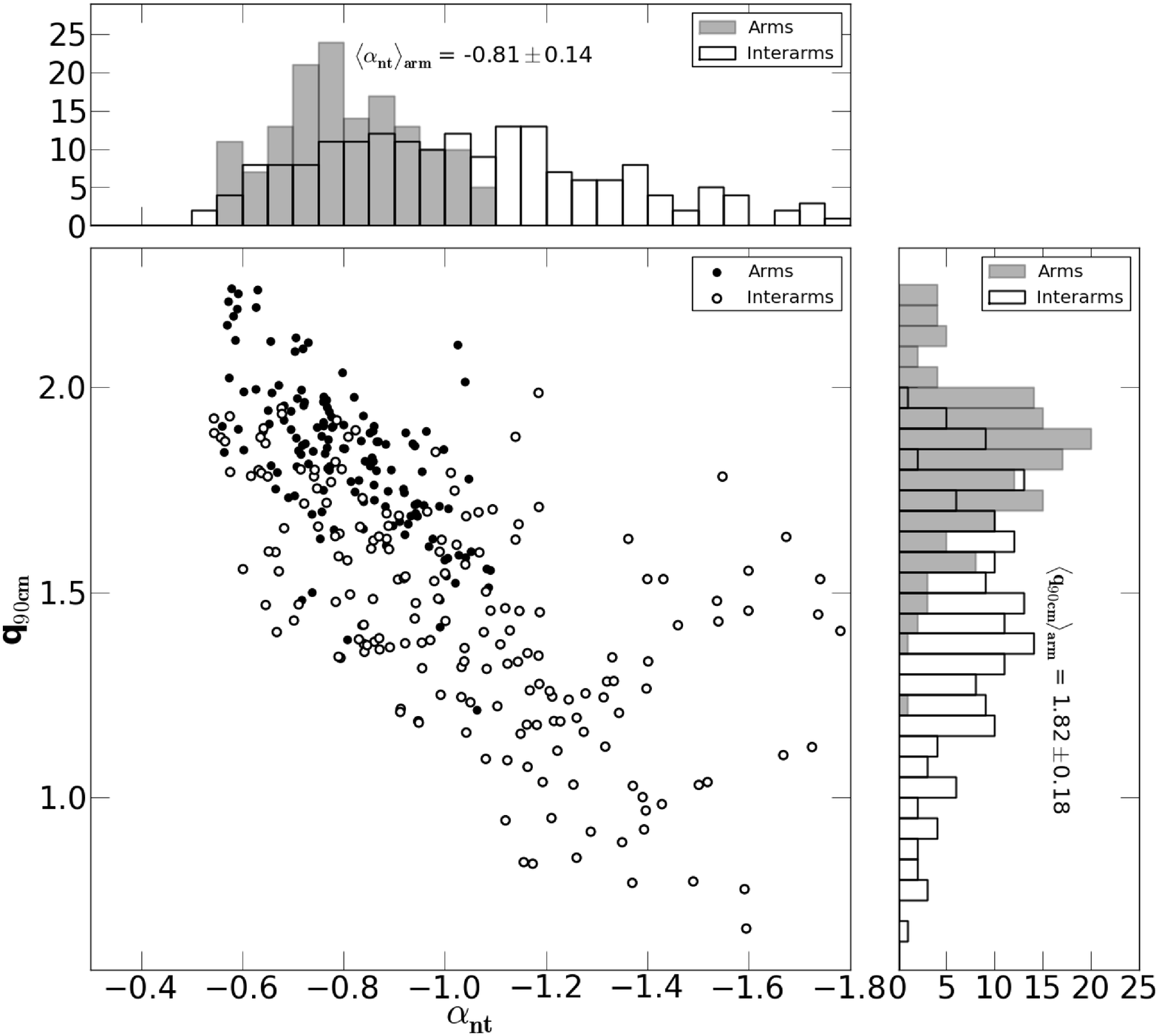}}} \\
\end{tabular}
\caption{The figure shows the distribution of $\ant$ with $q_{20\rm cm}$ (left)
and $q_{90\rm cm}$ (right).  The top panel shows the distribution of $\ant$
estimated from $\lambda90$cm and $\lambda20$cm nonthermal emission radio maps
within an area of 40$\times$40 arcsec$^2$ and a step size of 0.05, while the
right side horizontal panel shows the distribution of $q_\lambda$ within the
same area and bin size of 0.05.  The filled circles and histograms are for the
arms and unfilled circles and histograms are for interarm regions for all the 4
galaxies combined.}
\label{dualdistr}
\end{center}
\end{figure*}

The quantity `$q_\lambda$' can be expressed in terms of CRe escape timescale
($\tau_{\rm esc}$) and synchrotron timescale ($\tau_{\rm syn}$) as $q_\lambda
\propto \log (\tau_{\rm syn}/ \tau_{\rm esc})$ \citep[][]{murgi05, palad06}.
CRe emitting for these timescales also determines the variations in nonthermal
spectral index\footnote{In the text we define the spectral index $\alpha$ as,
$S_\nu \propto \nu^{\alpha}$} ($\ant$) from sites of injection to regions of
dominant energy loss. In Fig.~\ref{dualdistr} we also study the variation of
$\qlband$ (left) and $\qpband$ (right) with the $\ant$ determined at scales of
$\sim1-1.5$ kpc for all these galaxies from \citet{basu12}. The distribution of
$\ant$ with steps of 0.05, are shown in the top panels (Figure 10 in
\citealt{basu12}).  The right panels shows the distribution of $q_\lambda$ for
the respective frequencies with steps of 0.05.  The shaded histograms
represents the arms and the unfilled histograms are for the interarms. At
$\lambda20$ cm there is no apparent variation in the value of $\qlband$
with $\ant$ (Pearson's correlation coefficient, $r = 0.27$), while at
$\lambda90$ cm the $\qpband$ decreases with steepening of the $\ant$ (r =
0.7).

In the arm regions where $\ant$ is seen to have a narrow distribution with mean
$-0.8$ and dispersion of 0.14, the $q_\lambda$ values are higher, suggesting
$\tau_{\rm syn} \gg \tau_{\rm esc}$. However, in the interarm regions, $\ant$
and $q_\lambda$ have a wide distribution with more than 50 percent dispersion.
The values of `$q_\lambda$' also systematically decreases as one moves from
arms to interarms, indicating $\tau_{\rm esc} \gg \tau_{\rm syn}$ and thus the
CRe loose energy before escaping the disk giving rise to steeper $\ant$.
Similar results were found for IC 342 and NGC 5194 \citep{murgi05, palad06}.

\section{Discussion}
\label{discussion}

We have studied the radio--FIR correlation at $\sim$1 kpc scales for four
normal galaxies using nonthermal radio maps at $\lambda90$ cm and $\lambda20$
cm and the far infrared maps at $\lambda70~\mu$m. From the basic synchrotron
theory (e.g., \citealt{moffe75}) and considering the radio emission from CRe
emitting at critical frequencies, the energy of CRe at $\lambda90$ cm is
$\sim$1.5 GeV and at $\lambda20$ cm is $\sim$3 GeV when they are gyrating in a
typical magnetic field of $\sim10 ~\mu$G.  The far infrared emission at
$\lambda70~\mu$m originates from cool dust at $\sim$20 K heated by the
interstellar radiation field (ISRF) due to $\rm \sim5-20~M_\odot$ stars
\citep{dever89, xu90, xu96, dumas11}.  We separately examine these correlations
for the arm and the interarm regions, that is, regions of high and low thermal
fractions respectively.  The results of the various parameters as discussed in
Section~\ref{results} are given in Table 3 for individual galaxies, and here we
discuss the average properties.  The dispersion on the parameter $q_\lambda$ is
a measure of the tightness of the radio--FIR correlation, which for the arm
region is found to be less than 10 percent around the mean $q_\lambda$ for both
$\lambda20$ cm and $\lambda90$ cm.  For the interarm region the dispersion is
seen to increase to around 20 percent for both the frequencies.  Further we
find the slope of the radio--FIR correlation for the arm regions (also the high
thermal fraction regions) remains similar at both the radio frequencies (see
Table 3).  It should be noted that a large number of global scale
radio--FIR correlation studies exist, where the observed slope is steeper and
closer to unity \citep[see e.g.,][and the references therein]{price92, yun01}.
However, the spatially resolved studies relating FIR cool dust emission to
$\lambda$20 cm radio emission, yields a value of the slope $\sim0.6-0.9$ for
LMC \citep[]{hughe06} and $0.80\pm0.09$ for M31 \citep[]{hoern98}.  It is
difficult to compare the slopes obtained in global studies with the spatially
resolved case.  The flux in global studies are averaged over both arm and
interarm regions and we are uncertain about the contribution from each
component. Multifrequency spatially resolved studies can provide an
understanding of the relation between global scale and spatially resolved
studies. For the present case, in the interarm regions (regions of low
thermal fraction) for $\lambda20$ cm the slope is slightly flatter as compared
to the arms (see Eq. 1 and 2).  However, at $\lambda90$ cm, the slopes become
distinctly flatter than the arm regions (see Fig.~\ref{radio-fir} and Eq. 3 and
4).

Our results can be used to determine the coupling between magnetic field ($B$)
and the gas density ($\rho_{\rm gas}$) as discussed in the introduction and
thereby validating the `equipartition' assumptions in these galaxies at 1 kpc
scales.  \citet{dumas11} showed that the slope of the radio--FIR correlation
relates to $\kappa$ as, 
\begin{numcases}{\kappa =}
\frac{n~b}{3-\ant}, & optically thick dust \\
\frac{(n+1)~b}{3-\ant}, & optically thin dust
\end{numcases}
where, $n = 1.4$$\pm0.15$ is the Kennicutt-Schmidt law index \citep[see
e.g.,][]{kenni98}, $b$ is the slope of the radio--FIR correlation and $\ant$ is
the nonthermal spectral index. For these face-on galaxies we use the assumption
of optically thin dust to UV photons to estimate $\kappa$.  We find that
$\kappa = 0.51\pm0.1$ at $\lambda20$ cm and $\kappa = 0.4\pm0.1$ at $\lambda90$
cm. Similarly, for interarm regions due to a large range of $\ant$ we find
$\kappa$ in the range 0.41 -- 0.5 at $\lambda20$ cm and between 0.18 -- 0.22 at
$\lambda90$ cm. Our estimated values of $\kappa$, using the correlation between
$\lambda20$ cm and $\lambda70~\mu$m, are consistent with the predictions of
numerical MHD simulations of different ISM tubulence models, where $\kappa \sim
0.4-0.6$ \citep[see e.g.,][]{fiedl93,kim01, thomp06, grove03}.

In the arm regions, the slope and thus $\kappa$ remains similar for both
$\lambda20$ cm and $\lambda90$ cm.  Note that the above prescription to
determine $\kappa$ is valid provided the radio and the FIR emission arises from
the same emitting volume, with a diameter of about 1 kpc for most of the
observations reported here.  In the arm regions the UV photon has a mean free
path of $\sim$100 pc within which most of the FIR emission arises.  On the
other hand, the CRe which gives rise to the radio emission diffuse farther away
to $\sim1$ kpc at 1400 MHz and $\sim2$ kpc at 333 MHz in a galactic magnetic
field of $\sim10~\mu$G. Hence in order to have a similar slope with frequency,
the energy spectrum of the CRe giving rise to the radio emission should be
independent of the volume element. This can only happen if the timescale for
CRe diffusion/propagation ($\tau_{\rm diff}$) is significantly larger than
their generation timescale ($\tau_{\rm gen}$).  It turns out that the
$\tau_{\rm diff}$ is about $8\times10^7$ yr at 333 MHz and $4\times10^7$ years
at 1400 MHz which is significantly larger than the $\tau_{\rm gen}$  as evident
from the supernova rates, which is one every $10^4 - 10^5$ yr kpc$^{-2}$ in
Milky Way. We assume the same rate for these galaxies.

\begin{figure}
\begin{center}
\begin{tabular}{c}
\includegraphics[width=7cm]{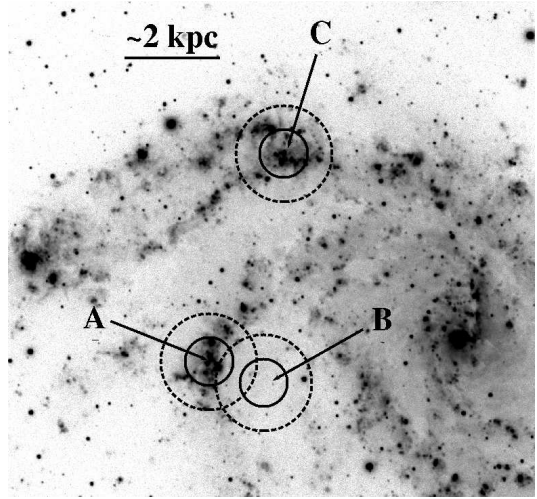}
\end{tabular}
\caption{The H$\alpha$ image of the galaxy NGC 6946 (KPNO 2-m telescope,
filter: KP1563) obtained from the ancillary data at SINGS website. The solid
circles represents the diffusion scale of $\sim$1 kpc for $\sim3$ GeV CRe at
$\lambda20$ cm, while the dashed circles represents the diffusion scales of
$\sim$2 kpc for $\sim$1.5 GeV CRe at $\lambda90$ cm. See text for the details.}
\label{spiral}
\end{center}
\end{figure}

The slope of the radio--FIR correlation in the interarm (low thermal fraction)
region is similar to that of the arm at $\lambda20$ cm, however it becomes
distinctly flatter at $\lambda90$ cm. The flattening primarily happens due to
relative increase in radio flux at $\lambda90$ cm as compared to $\lambda20$
cm, which has the effect that $\ant$ gradually becomes steeper in the
interarms. This relative increase in $\lambda90$ cm flux can be explained  by
continuous generation of CRe in the arm, which subsequently propagates into the
interarm (e.g., from A to B or from farther regions in arms like C to B in
Fig.~\ref{spiral}).  The propagation timescale for these CRe are few times
$10^7$ years assuming Alfv$\acute{\rm e}$n velocity of 100 km s$^{-1}$ and
typical arm to interarm distance of 1--2 kpc.  In such a scenario, using
Equation 6 of \citep{karda62}, in a typical galactic magnetic field of
$\sim10~\mu$G, there would be a break in the energy spectrum for electrons
above $\sim$2 GeV.  This break frequency lies below $\lambda90$ cm or above 333
MHz.  Such breaks have been seen at $\sim900$ MHz and $\sim$1 GHz for similar
normal galaxies, NGC 3627 and NGC 7331 respectively \citep{palad09}.  Thus the
CRe emitting at $\lambda90$ cm, which lie above the break, do not loose
significant amount of energy as compared to their higher energy counterparts.
Hence, this results in increasing the relative flux at $\lambda90$ cm. 

For the slope to remain similar between arms and interarm regions at
$\lambda20$ cm (below the break), the ratio of the radio to FIR flux densities
should remain similar. Observed radio flux between arm and interarm changes by
a factor of $\sim$2--2.5.  Similar ratio of flux density between arm and interarm
regions at $\lambda20$ cm can be caused due to steeping of the spectral index
to $\lesssim -1.1$ as compared $\sim-0.6 ~\rm to -0.8$ in the arms.  This
implies the FIR flux should change by a factor of $\sim$2.5--3 between arm and
interarm regions for radio--FIR slope of $\sim$0.8.  The FIR flux density
($F_\lambda$) depends on the dust temperature ($T_{\rm dust}$) and its density
($\rho_{\rm dust}$) as, $F_\lambda \propto \rho_{\rm dust} Q_{\rm abs}(a,
\lambda) B_\lambda(T_{\rm dust})$, where $Q_{\rm abs}(a, \lambda)$ is the FIR
wavelength ($\lambda$) dependent absorption coefficient for gain radius, $a$
\citep{drain84, alton04}.  The temperature do not change significantly between
arm and interarm for these galaxies \citep{basu12}. For a constant gas-to-dust
ratio, i.e, $\rho_{\rm dust} \propto \rho_{\rm gas}$, a factor of 2--4 drop in
average gas density between arm and interarm regions \citep[found using the
CO$_{\rm J:2\to1}$ maps from Heracles;][]{leroy09} would therefore cause the
factor of 2--3 drop in FIR emission.

The slope of 0.8$\pm$0.1 of the radio--FIR correlation indicates that the
energy equipartition assumption between cosmic ray particles and magnetic
field may be valid in the gas rich arms of the galaxies at our spatial
resolution of $\sim$1 kpc.  For the interarm regions at $\lambda20$ cm the
slope is similar to what is seen in arms, and thereby satisfying the
equipartition conditions. The flattening of the slope at $\lambda90$ cm does
not indicate any break down of equipartition condition, but results due to
overlapping emissions from adjacent regions.


\acknowledgments  We thank Adam Leroy for kindly providing us the FITS files
for the CO$_{\rm J:2\to1}$ moment-0 maps. We thank Yogesh Wadadekar for
useful comments.  We also thank the anonymous referee for valuable comments.
This research has made use of the NASA/IPAC Extragalactic Database (NED), which
is operated by the Jet Propulsion Laboratory, California Institute of
Technology, under contract with the National Aeronautics and Space
Administration.  This work is based (in part) on observations made with the
{\it Spitzer Space Telescope}, which is operated by the Jet Propulsion
Laboratory, California Institute of Technology under a contract with NASA.
 




\end{document}